\documentclass[a4paper,11pt]{article}
\usepackage{pos}
\usepackage{lineno}
\usepackage{pos}
\usepackage{subfig}
\usepackage{float}
\usepackage{xcolor}
\usepackage{booktabs}
\usepackage{multirow}
\usepackage{setspace}

\title{Optimization of the optical array geometry for IceCube-Gen2}
 \ShortTitle{IceCube-Gen2 geometry studies}

\author{The IceCube-Gen2 Collaboration \\{\normalsize \normalfont(a complete list of authors can be found at the end of the proceedings)}}

\emailAdd{anastasiia.yovych@icecube.wisc.edu}

\abstract{IceCube-Gen2 is a planned extension of the IceCube Neutrino Observatory at the South Pole designed to study the high-energy neutrino sky from TeV to EeV energies with a five times better point source sensitivity than the current IceCube detector. This is achieved by deploying 120 new strings with attached optical sensors in a pattern around IceCube that features considerably larger distances between individual strings than the $\sim$125\,m for the existing detector. Here, we present the results of an optimization study searching for the best point source sensitivity while varying the IceCube-Gen2 string spacing between 150\,m and 350\,m. \\

\vspace{4mm}
{\bfseries Corresponding author:}
Anastasiia Omeliukh$^{ a, *}$\\
{$^{a}$ \itshape DESY, Platanenallee 6, 15738 Zeuthen, Germany}\\
[4mm]
$^*$ Presenter

\FullConference{37$^{\rm{th}}$ International Cosmic Ray Conference (ICRC 2021)\\
		July 12th -- 23rd, 2021\\
		Online -- Berlin, Germany}

}

\begin{document}
\maketitle

\section{IceCube-Gen2}
IceCube is a cubic-kilometer neutrino detector installed in the ice at the geographic South Pole~\cite{Aartsen:2016nxy} between depths of 1450\,m and 2450\,m, completed in 2010. Reconstruction of the direction, energy and flavor of the neutrinos relies on the optical detection of Cherenkov radiation induced by charged particles produced in the interactions of neutrinos in the surrounding ice or the nearby bedrock.

IceCube-Gen2 will be an extension of the existing IceCube \cite{Aartsen_2021} array aiming to explore the high-energy neutrino sky. 120 new strings with attached optical sensors will be installed around IceCube with considerably larger distances between individual strings, increasing the sensitivity for neutrino energies above 10~TeV.  A large surface array will enhance the veto of cosmic-ray (CR) air showers and, thus, lower the energy threshold for the identification of neutrino interactions on the Southern sky. The detection of neutrinos with energies above 100~PeV is the main goal of the radio array \cite{Allison_2015}. As a first step towards IceCube-Gen2, IceCube Upgrade~\cite{Ishihara:2019aao}, currently under construction,  lowers the detection threshold for neutrinos to 1~GeV leading to significant improvements for oscillation measurements, dark matter, and other beyond Standard Model physics searches. Here we focus on the optimization of the inter-string spacing for best point source sensitivity of the high-energy optical array that is comprised of the 86 IceCube and 120 additional strings.

The IceCube-Gen2 optical array will, like IceCube, detect neutrinos via the relativistic products of charge current (CC) and neutral current (NC) interactions. Muons created in $\nu_\mu$~CC interactions produce tracks when passing through the detector. Cascades are produced for all $\nu_e$ and $\nu_\tau$, as well as $\nu_\mu$ NC interactions. In this work, we study the sensitivity to point sources with tracks.

\section{``Sunflower'' geometries}

\begin{figure}[h]
    \centering
   
    \subfloat[\centering s=150 m]{{\includegraphics[width=4.7cm]{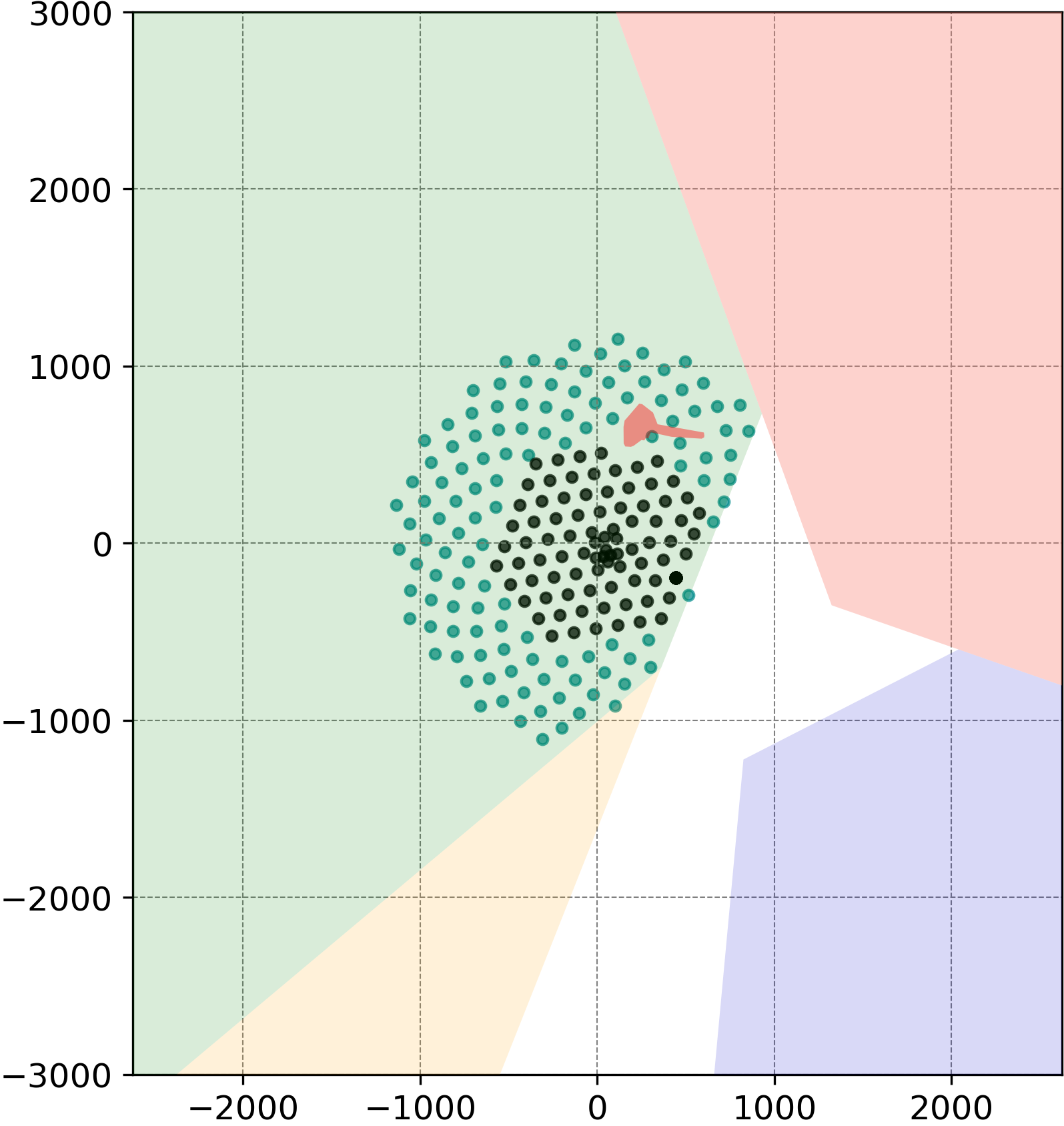} }}%
    \hspace{0.5em}
    \subfloat[\centering s = 240 m]{{\includegraphics[width=4.7cm]{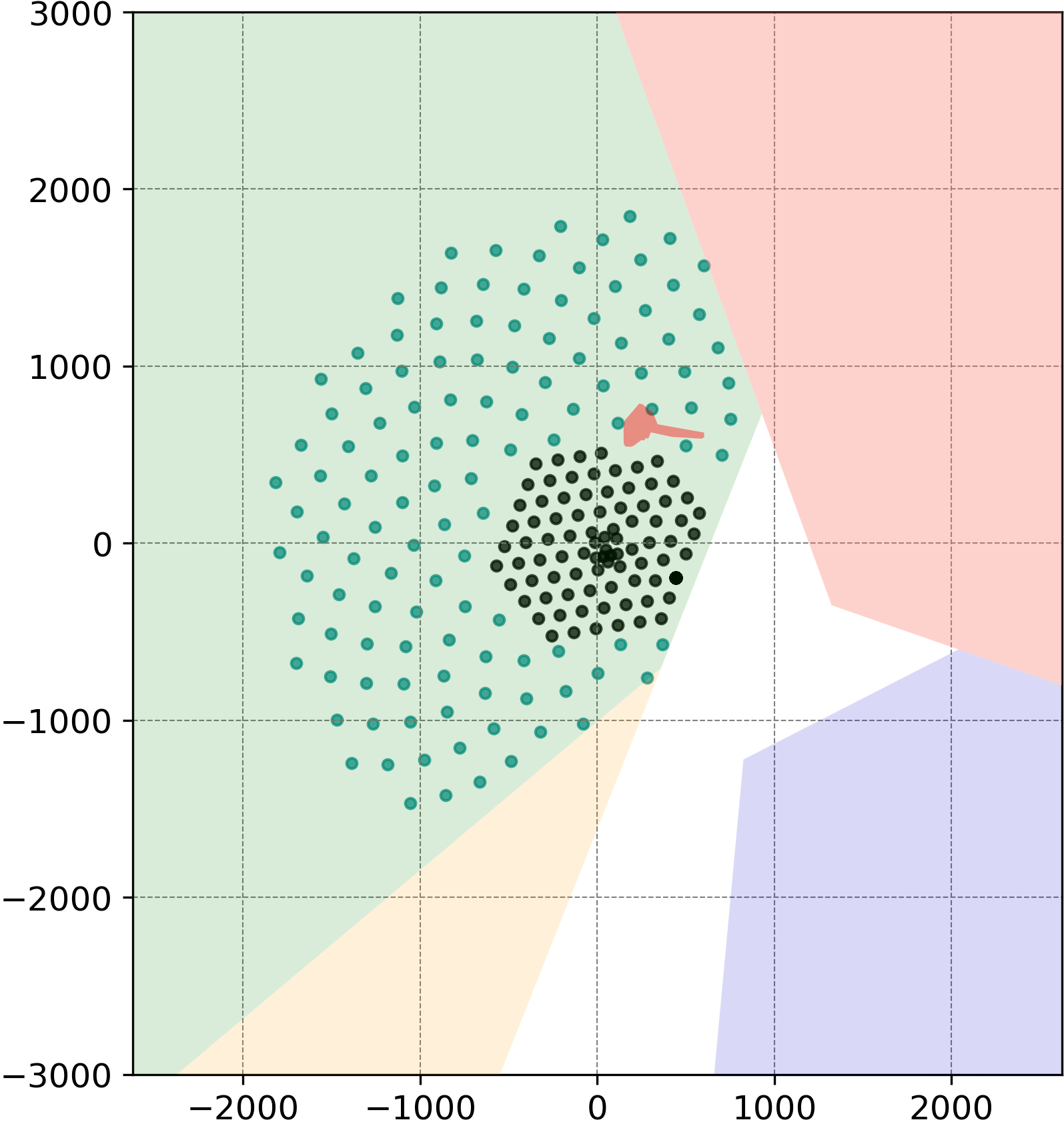} }}%
   \hspace{0.5em}
    \subfloat[\centering s = 350 m]{{\includegraphics[width=4.7cm]{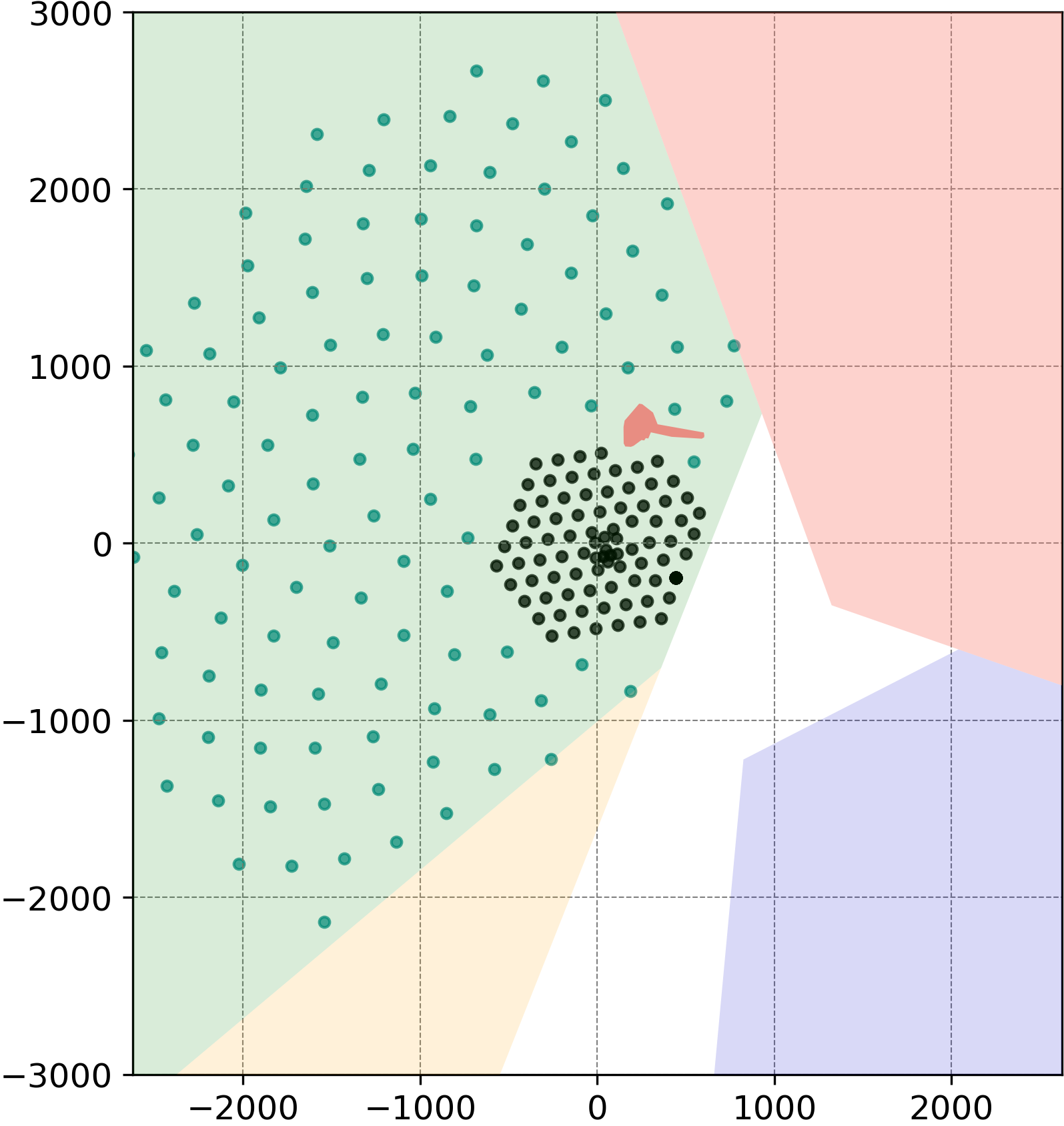} }}%
    
    \caption{Visualization of IceCube-Gen2 geometries with inter-string spacings of 150 m (a), 240 m (b), and 350 m (c). The black dots represent the location of existing IceCube strings. The small red region near the IceCube detector is an old South Pole station and unsuitable for deployment. The allowed region for IceCube-Gen2 construction is limited to the light-green sector. All other sectors are reserved for different purposes.}%
    \label{fig:geo-comparison}%
\end{figure}

A previous study of geometrical shapes of the IceCube-Gen2 optical array \cite{Lunemann:2017Ji} has shown that a ``Sunflower''-like geometry is advantageous compared to IceCube's regular grid. The location of the strings in ``Sunflower'' geometry is defined in a polar coordinate system as
\begin{equation} \label{eq1}
    r = s \sqrt{n}, \; \; \phi = \frac{2\pi}{g^2} n
\end{equation}
where $g=\frac{1+\sqrt{5}}{2}$ (golden ratio), $s$ is a spacing parameter, and $n$ is a natural number that defines string position.
To generate the ``Sunflower'' geometry, the spiral starts at the center of  IceCube. All the string positions that are inside of IceCube region or outside of the allowed (cf. Fig. \ref{fig:geo-comparison}) construction region are removed. The process is stopped when the total number of 120 strings is reached.

The IceCube-Gen2 optical sensors will be placed between 1325~m and 2575~m underground with 16~m spacing between the DOMs along the string. Eight values of inter-string spacing parameter $s$ were chosen for the optimization study: 150\,m, 200\,m, 220\,m, 240\,m, 260\,m, 280\,m, 300\,m, and 350\,m. Fig.\ref{fig:geo-comparison} shows the schematic top view of the high-energy optical array layout for three of these parameters (150\,m, 240\,m, and 350\,m).

\section{Monte Carlo simulations}

IceCube-Gen2 will have novel optical sensors such as mDOMs and D-Eggs \cite{Aartsen_2021}. The multi-PMT DOM (mDOM) is an optical module with 24 3-inch photo-multiplayers (PMTs). This type of DOM will have 2.2 times higher photocathode area in comparison with IceCube pDOM, isotropic sensitivity, and opportunity to obtain information from individual ``pixels'' (24 PMTs). To be able to use already developed and well-tested techniques to reconstruct simulated events with new sensors, mDOMs were simulated as pDOM type modules, installed on existing IceCube strings, that have $\sim$3 times higher quantum efficiency and isotropic angular acceptance.

The IceCube-Gen2 point source sensitivity is dominated by the detection and reconstruction capabilities for tracks produced in $\nu_\mu$ CC interactions due to larger effective area and better angular resolution compared to cascades. Therefore,  for this study, an isotropic distribution of muons having a power-law energy spectrum with a spectral index $\gamma=-1.4$ in the energy range between $E_{min}$ = 300~GeV and $E_{max}$~=~100~PeV is generated. The spectral index is chosen to keep the spectrum approximately flat in logarithmic scale after the event selection.  Individual stochastic muon energy deposits are simulated inside of a cylindrical region that encloses IceCube-Gen2 with the height $H=1700$\,m and the radius $R=2500$\,m. In the next step, the CLSim package is used -- a simulation software package to propagate photons from Cherenkov radiation \cite{9041727}. The final simulation step is the detector simulation that simulates noise and the charge response of the PMTs. 

The properties of the simulated tracks are reconstructed using IceCube's reconstruction algorithms. The best performing muon track direction reconstruction algorithm (SplineMPE) is based on a maximum likelihood method using the arrival time distribution of Cherenkov photons registered by the experiment’s PMTs \cite{abbasi2021muontrack}. It uses the results of simpler reconstruction algorithms, such as the LineFit algorithm \cite{Stenger1990}, as starting hypothesis. 

Eight datasets (one per geometry) were produced containing $\sim$ 50 000 triggered events in each.

\section{Event selection}
An event selection is needed to exclude poorly reconstructed tracks from the dataset. Four parameters are used to discriminate well-reconstructed tracks from poorly-reconstructed tracks:
\begin{itemize}
    \item \textbf{LineFit velocity} is a parameter describing an apparent track velocity in the LineFit algorithm. As a ``sanity cut'', this value should be less than twice the speed of light in ice. The major part of IceCube events have LineFit velocity less than speed of light in ice.
    
    \item \textbf{Reduced log-likelihood} is the absolute value of the best-fit log-likelihood divided by the number of degrees of freedom which is analogous to $\chi^2/n_{dof}$. ``Good'' tracks in IceCube have a value of less than 8.5 in this variable.
    \item \textbf{Number of DOMs} with ``direct'' photons, i.e. that arrive between -15 ns and +50 ns of the expected arrival time in a scatter-free medium, given the best-fit track. Good quality tracks in IceCube have a value larger than 6 for this parameter.
    \item \textbf{Track length} within the instrumented volume of IceCube. The direction of tracks that intersect only a small portion of the instrumented volume can usually not be well reconstructed. Therefore, the reconstructed track length must be greater than 120 m in IceCube. 
\end{itemize}

\begin{figure}
    \centering
    \includegraphics[width=14cm]{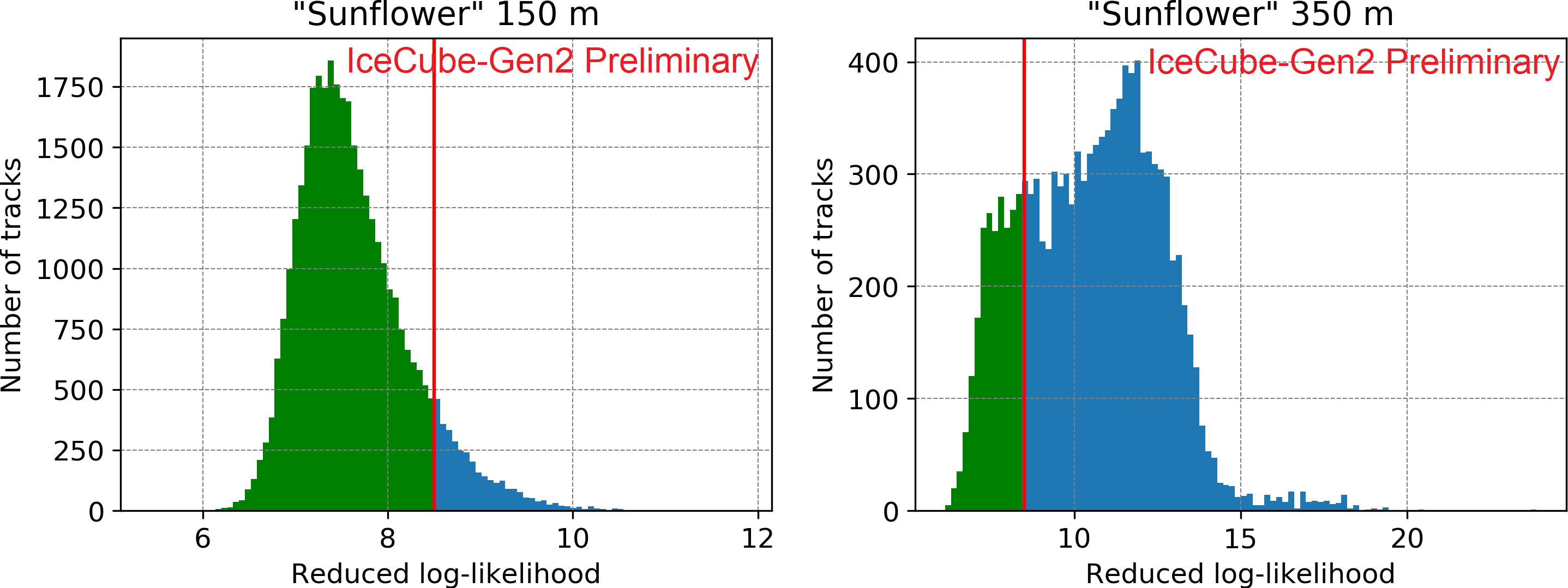}
    \caption{Distribution of  the reduced log-likelihood parameter for two different high-energy array geometries. The part of dataset remaining when applying the (unaltered) cut motivated from IceCube analysis is marked green. 
    }
    \label{fig:rlogl_cut}
\end{figure}

The IceCube quality cuts might be inapplicable for IceCube-Gen2, therefore the performance of this event selection was examined for Gen2. The distributions of LineFit velocity, number of DOMs with direct photons, and track length did not change with inclusion of the high-energy array, unlike the reduced log-likelihood, which heavily depends on the spacing parameter.  This is demonstrated also in Fig.\,\ref{fig:rlogl_cut}: The part of dataset that remains after the standard IceCube selection procedure is marked green (on left from the red line) . The high cut efficiency for IceCube (125\,m inner-string spacing) or ``Sunflower'' 150 m drops dramatically for the Sunflower with 350 m inter-string spacing. Therefore the reduced log-likelihood parameter had to be re-optimized individually for each geometry for this study.

Two alternate sets of cuts were defined monitoring the selection efficiency in two different energy ranges. A high-energy (HE) set, aiming for a high efficiency for muons with $E_\mu > 100$~TeV, and a low-energy (LE) set, aiming for a high-efficiency for muons with $E_\mu < 50$ TeV.
As the former selection is more stringent than the latter, the LE selection implicitly features a high efficiency also for high-energy tracks. The efficiency is checked separately in three zenith bins. The main criterion for defining the cut values was to maintain a small median opening angle between true and reconstructed direction and have simultaneously a good cut efficiency. In addition, the 99\% quantile of the point spread function was considered as an estimate for the size of the tails of poorly reconstructed events in the set. Lacking a dedicated simulation of atmospheric muon background, the 99\% quantile of the PSF has been used as a proxy for the angular reconstruction performance on this background. The selection cut has been adjusted to ensure that $\geq$ 99\% of the muons arriving from zenith angles $\leq$ 80 degrees are reconstructed outside the region with zenith angles $\geq$ 90$^\circ$. The values of the selection cuts, as well as median opening angle, 99\% quantile for the point spread function (PSF), and cut efficiency are shown in Table \ref{tab:table1} and Table \ref{tab:table2} for the high- and low-energy sample, respectively. For horizontal muons, the angular resolution is approximately the same for all geometries. However, in zenith bin $50\pm10^\circ$, the denser instrumentation will show significantly better angular resolution.

\setlength{\aboverulesep}{-0.5pt}
\setlength{\belowrulesep}{-0.5pt}

\begin{table}
\centering
\begin{tabular}{ c c |c c c| c c c| c c c }
  \toprule
    \multirow{2}{*}{s, m}& \multirow{2}{*}{`rlogl' cut} & \multicolumn{3}{c}{Zenith $90\pm 10^\circ$} & \multicolumn{3}{|c}{Zenith $70\pm 10^\circ$} & \multicolumn{3}{|c}{Zenith $50\pm 10^\circ$}\\
    \cmidrule{3-11}
    & & $\eta$ & $\Psi_{50}$ & $\Psi_{99}$ & $\eta$ & $\Psi_{50}$ & $\Psi_{99}$  & $\eta$ & $\Psi_{50}$ & $\Psi_{99}$  \\
    \hline
    150 & 8.5 & 97\% &  0.23 & 3.8 & 93\% &  0.27 & 10.9  & 90\%  & 0.28 & 14.3 \\
    200 & 9.0 & 96\% & 0.21 & 5.0 &  93\% &  0.33 & 11.8 & 89\%  & 0.36 & 14.8\\
    220 & 9.0 & 97\%  & 0.22 & 6.1 & 97\% &  0.33  & 11.2 & 93\%  & 0.40 & 31.4\\
    240 & 9.5 & 98\% &  0.23 & 4.9 & 98\% &  0.40 & 17.4 & 96\%  & 0.43 & 13.3\\
    260 & 9.5 & 95\% &  0.22 & 8.5 & 94\% &  0.30 & 10.3 & 89\%  & 0.41 & 18.9 \\
    280 & 10.0 & 96\%  & 0.20 & 5.2 & 92\% & 0.35 &  8.1 & 86\%  & 0.49 & 12.7\\
    300 & 10.5 & 93\% & 0.20 & 8.3 & 89\% & 0.31 & 8.2&   87\%  & 0.49 & 18.7\\
    350 & 11.5 & 95\% &  0.21 &  7.7 & 78\% & 0.34 & 7.6 &  72\%  & 0.57 & 20.7\\
  \bottomrule
\end{tabular}%
\caption{\label{tab:table1} Reduced log-likelihood cuts optimized for the high-energy muon sample ($E_\mu > 100$ TeV). $\eta$ denotes cut efficiency on this sample (i.e. fraction of events that remain after the application of the cut), $\Psi_{50}$ is the median opening angle (degrees), and $\Psi_{99}$ is the 99\% quantile of the opening angle (degrees).}
\end{table}

\begin{table}[h]
\centering
\begin{tabular}{ c c |c c c| c c c| c c c }
  \toprule
    \multirow{2}{*}{s, m}& \multirow{2}{*}{`rlogl' cut} & \multicolumn{3}{c}{Zenith $90\pm 10^\circ$} & \multicolumn{3}{|c}{Zenith $70\pm 10^\circ$} & \multicolumn{3}{|c}{Zenith $50\pm 10^\circ$}\\
    \cmidrule{3-11}
    & & $\eta$ & $\Psi_{50}$ & $\Psi_{99}$ & $\eta$ & $\Psi_{50}$ & $\Psi_{99}$  & $\eta$ & $\Psi_{50}$ & $\Psi_{99}$  \\
    \hline
    150  & 8.5 & 91\% &  0.24 & 5.4 & 90\%  & 0.34 & 6.9 & 87\%  & 0.44 & 10.8 \\
    200 & 9.5 & 89\% &  0.26 & 7.2 &  88\%  & 0.37 & 9.5 & 85\%  & 0.48 & 15.3\\
    220 & 10.0 & 88\% & 0.25 & 6.9    & 88\%  & 0.36 & 10.1 &  85\% & 0.51 & 12.9 \\
    240 & 10.5& 90\% &  0.25& 6.6 & 89\% & 0.38 & 9.5  & 87\% & 0.54 & 16.9  \\
    260 & 10.5& 83\% &  0.25 & 9.2   & 82\%  & 0.38 & 13.9 & 81\%  & 0.55 & 15.2\\
    280 & 11.0 & 84\% & 0.26 & 7.7   & 84\%  & 0.40 & 8.6 & 79\%  & 0.53 & 20.2\\
    300 & 11.5 & 84\%  & 0.26 & 8.5 & 84\% & 0.40 & 13.0 & 78\% & 0.59 & 15.8 \\
    350  & 12.5 & 83\%  & 0.27 & 6.6 &  78\% & 0.40 & 13.9  & 77\%  & 0.58 & 21.0\\
  \bottomrule
\end{tabular}%
\caption{\label{tab:table2} Reduced log-likelihood cuts optimized for the low-energy muon subsample ($E_\mu < 50$ TeV). $\eta$ denotes cut efficiency on this sample (i.e. fraction of events that remain after the application of the cut), $\Psi_{50}$ is the median opening angle (degrees), and $\Psi_{99}$ is the 99\% quantile of the opening angle (degrees).}
\end{table}

\section{Sensitivity to astrophysical neutrinos}
In  the  following,  we use the method  described in \cite{vanSanten:2017chb} to obtain expected neutrino  event  rates and point source sensitivities from  basic detector performance quantities such as the muon effective area and the energy resolution.

The performance of each of the eight proposed detector geometries was parametrized. For muon tracks, the detector performance is characterized by four quantities: the muon effective area, the selection efficiency, the energy resolution, and the point spread function.

\begin{figure}
    \centering
    \includegraphics[width=8cm]{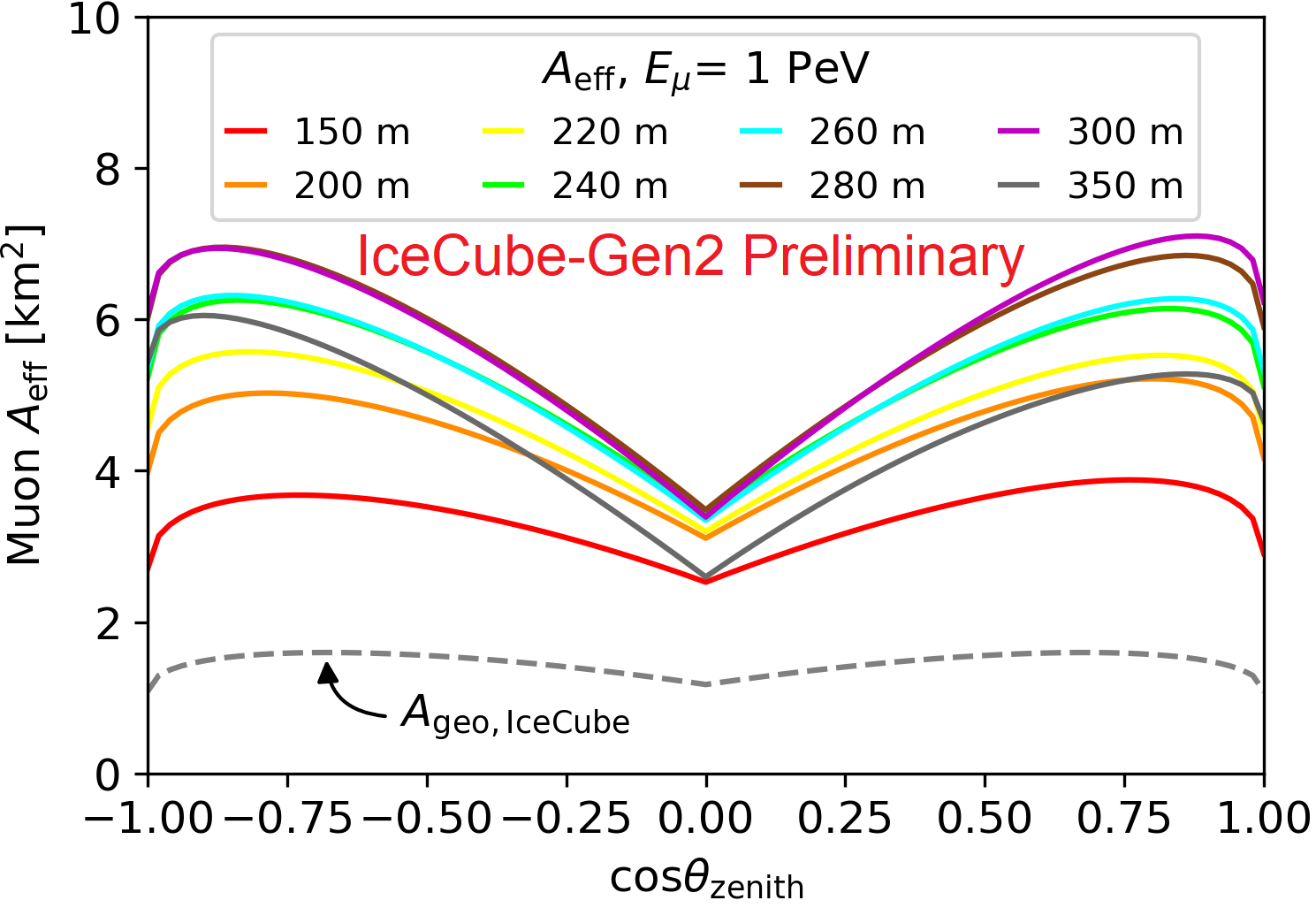}
    \caption{The average muon effective area after quality cuts as a function of zenith angle is shown for 1 PeV muons.}
    \label{fig:aeff}
\end{figure}

\textbf{Muon effective area}. Fig. \ref{fig:aeff} shows the muon effective area of the “Sunflower” geometries with spacings from 150 m to 350 m for muons entering the detector with 1 PeV energy. 

\textbf{Muon energy resolution}. 
The stochastic energy losses of highly energetic muons cause a significant spread in observed muon energy. The muon energy resolution is a parameterization of the reconstructed energy distribution for muons entering the detector with a specific energy, after the selection described above.

\textbf{Muon selection efficiency} is a ratio of the number of tracks that passed selection cuts to the full number of triggered tracks.

\textbf{Point spread function} parameterizes the distribution of the angular distance between the true muon direction and the reconstructed direction as a function of zenith angle and muon energy.

\section{Point source discovery potential}
The 5$\sigma$ discovery potential for point sources is used as measure to compare the sensitivity of different inter-string spacings. It is defined as the neutrino flux from a candidate neutrino source that is required to produce an excess of neutrinos over background equivalent to a significance of 5$\sigma$ in half of all simulated experiments.

While IceCube measurements have confirmed the existence of an astrophysical neutrino flux in multiple energy bands and for different neutrino flavors \cite{Niederhausen:2017mjk, abbasi2020icecube, Stettner:2019TL}, the observation of individual sources is limited to only few candidates \cite{2018, stein2021tidal}. However, the significantly larger instrumented area of IceCube-Gen2 will allow detection of fluxes from individual sources that are much fainter than current limits.

\begin{figure}
    \centering
    \subfloat[\centering HE cuts set for event selection]{{\includegraphics[width=7.2cm]{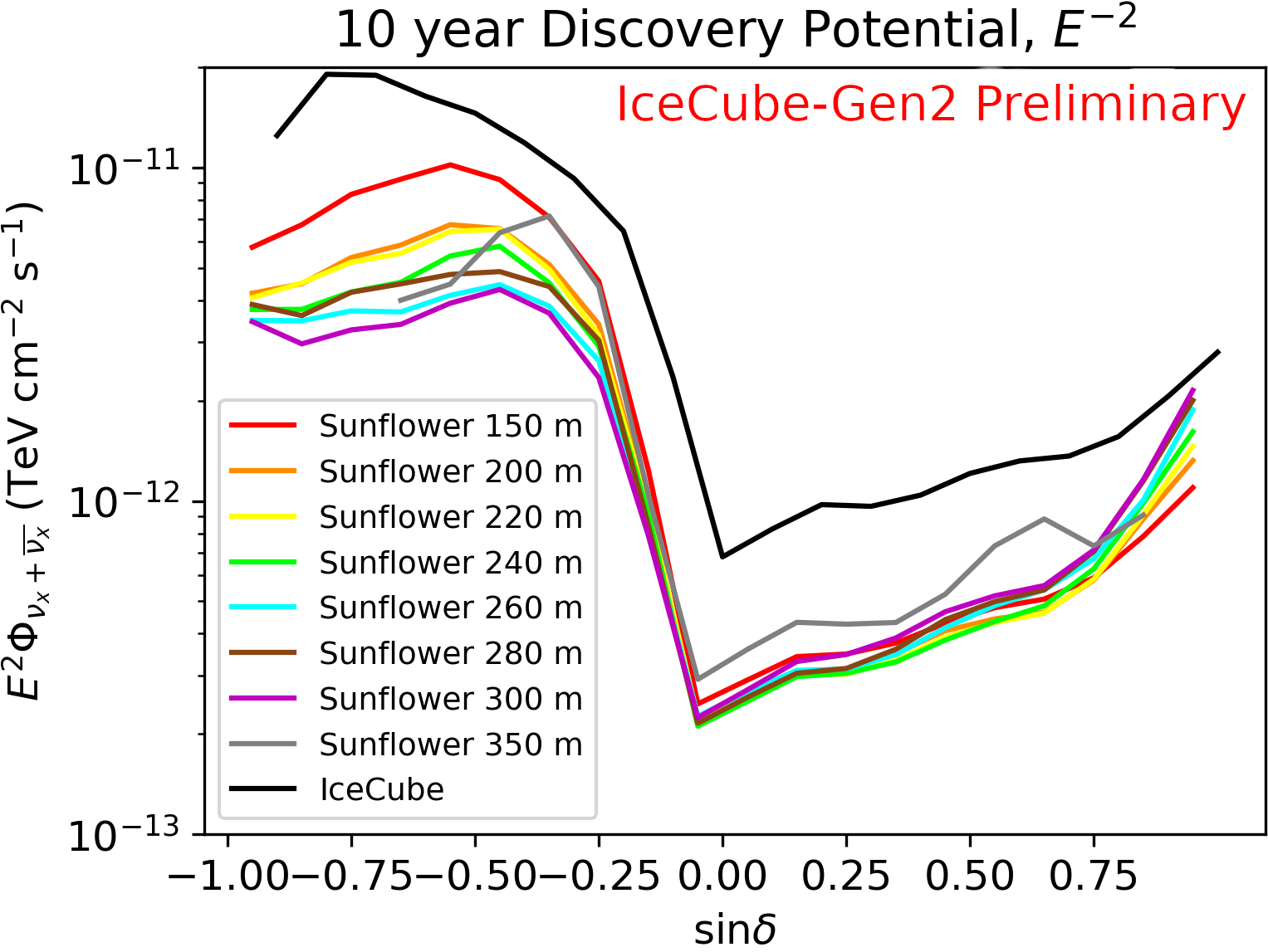} }}%
    \hspace{0.5em}
    \subfloat[\centering  LE cuts set event selection]{{\includegraphics[width=7.2cm]{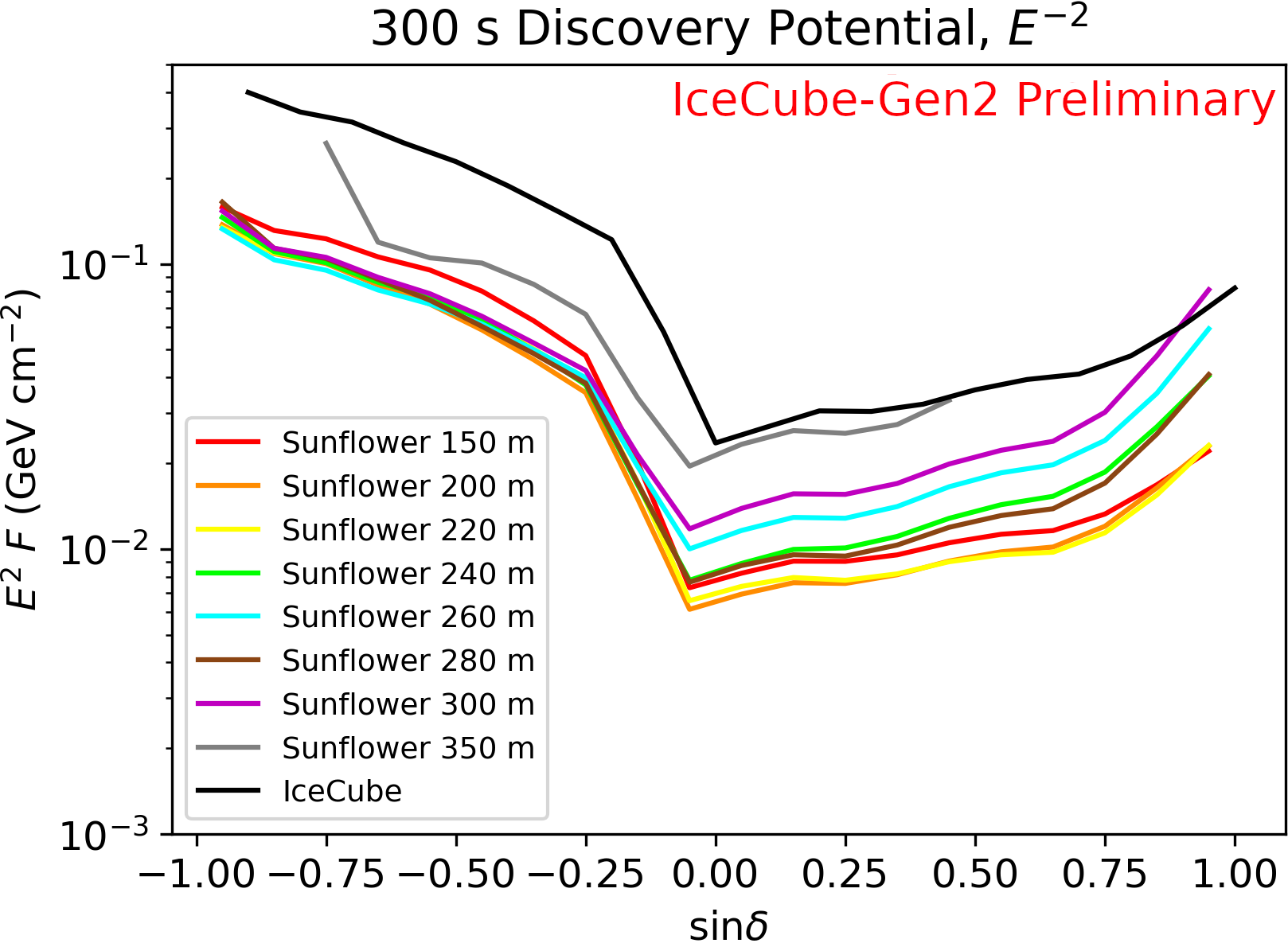} }}%
        \caption{IceCube-Gen2 (optical array, ``Sunflower'' geometries) discovery potential for (a) 10 years of exposure, high-energy cuts applied, (b) 300 seconds of exposure, low-energy cuts applied. }
    \label{im:dp}
\end{figure}

Figures  \ref{im:dp} (a) and \ref{im:dp} (b) show the discovery potentials to an $E^{-2}$ flux from a single source over 10 years and 300 seconds of exposure time, respectively, using only through-going tracks made by muons formed around IceCube and passing through the detector. The atmospheric background is assumed to be isotropic within one declination bin. For long exposure, the values of discovery potentials for the geometries with inter-spacing parameters from 200 m to 300 m differ only slightly (especially at the horizon where these values coincide within 5\% margin). ``Sunflower'' 350 m and ``Sunflower'' 150 m would have slightly worse sensitivity at the horizon compared to other geometries. For the ``burst'' discovery potentials (short signal duration), the smaller spacing values lead to better discovery potential, although the difference between the values of discovery potential is relatively small, as in case of 10 years of exposure.

The sensitivity in the direction of the Northern celestial pole is lower, because a significant fraction of high-energy neutrinos is absorbed in the Earth. In the Southern sky, the overall sensitivity is reduced due to the energy threshold imposed by the rejection of CR muon background, and the limited target mass for neutrino interactions between the surface and the detector. The largest projected density of instrumentation, the available target material and the large possible range of neutrino-induced muons make the celestial horizon ($\delta =0$) the region with the best sensitivity for both, IceCube and IceCube-Gen2.  

\section{Summary}

IceCube-Gen2 will be a multi-component facility that targets the detection of neutrinos with energies from several GeV to several EeV. Here we have presented sensitivity studies for the high-energy optical array by varying the inter-string spacing. Good quality tracks have been selected to determine the relation between point source discovery potential and inter-string spacing. Two selections have been defined, one that focuses on high efficiency for tracks with energies below 50~TeV and one that focuses on high efficiency for tracks above 100~TeV. For sources with a generic power-law spectrum with index $-$2, inter-string spacings between 200~m and 280~m show a similar performance with both event selections, while for larger and smaller inter-string spacings the discovery potential is comparatively worse.


\bibliographystyle{ICRC}
\bibliography{ref}


%
%
%




\clearpage
\section*{Full Author List: IceCube-Gen2 Collaboration}


\scriptsize
\noindent
R. Abbasi$^{17}$,
M. Ackermann$^{71}$,
J. Adams$^{22}$,
J. A. Aguilar$^{12}$,
M. Ahlers$^{26}$,
M. Ahrens$^{60}$,
C. Alispach$^{32}$,
P. Allison$^{24,\: 25}$,
A. A. Alves Jr.$^{35}$,
N. M. Amin$^{50}$
% ...
\\
\\
% ...
$^{12}$ Universit{\'e} Libre de Bruxelles, Science Faculty CP230, B-1050 Brussels, Belgium \\
% ...
$^{17}$ Department of Physics, Loyola University Chicago, Chicago, IL 60660, USA \\
% ...
$^{22}$ Dept. of Physics and Astronomy, University of Canterbury, Private Bag 4800, Christchurch, New Zealand \\
% ...
$^{24}$ Dept. of Astronomy, Ohio State University, Columbus, OH 43210, USA \\
$^{25}$ Dept. of Physics and Center for Cosmology and Astro-Particle Physics, Ohio State University, Columbus, OH 43210, USA \\
$^{26}$ Niels Bohr Institute, University of Copenhagen, DK-2100 Copenhagen, Denmark \\
% ...
$^{32}$ D{\'e}partement de physique nucl{\'e}aire et corpusculaire, Universit{\'e} de Gen{\`e}ve, CH-1211 Gen{\`e}ve, Switzerland \\
% ...
$^{35}$ Karlsruhe Institute of Technology, Institute for Astroparticle Physics, D-76021 Karlsruhe, Germany  \\
% ...
$^{50}$ Bartol Research Institute and Dept. of Physics and Astronomy, University of Delaware, Newark, DE 19716, USA \\
% ...
$^{60}$ Oskar Klein Centre and Dept. of Physics, Stockholm University, SE-10691 Stockholm, Sweden \\
% ...
$^{71}$ DESY, D-15738 Zeuthen, Germany \\
% ...

\subsection*{Acknowledgements}

\noindent
USA {\textendash} U.S. National Science Foundation-Office of Polar Programs,
U.S. National Science Foundation-Physics Division,
U.S. National Science Foundation-EPSCoR,
Wisconsin Alumni Research Foundation,
Center for High Throughput Computing (CHTC) at the University of Wisconsin{\textendash}Madison,
Open Science Grid (OSG),
Extreme Science and Engineering Discovery Environment (XSEDE),
Frontera computing project at the Texas Advanced Computing Center,
U.S. Department of Energy-National Energy Research Scientific Computing Center,
Particle astrophysics research computing center at the University of Maryland,
Institute for Cyber-Enabled Research at Michigan State University,
and Astroparticle physics computational facility at Marquette University;
Belgium {\textendash} Funds for Scientific Research (FRS-FNRS and FWO),
FWO Odysseus and Big Science programmes,
and Belgian Federal Science Policy Office (Belspo);
Germany {\textendash} Bundesministerium f{\"u}r Bildung und Forschung (BMBF),
Deutsche Forschungsgemeinschaft (DFG),
Helmholtz Alliance for Astroparticle Physics (HAP),
Initiative and Networking Fund of the Helmholtz Association,
Deutsches Elektronen Synchrotron (DESY),
and High Performance Computing cluster of the RWTH Aachen;
Sweden {\textendash} Swedish Research Council,
Swedish Polar Research Secretariat,
Swedish National Infrastructure for Computing (SNIC),
and Knut and Alice Wallenberg Foundation;
Australia {\textendash} Australian Research Council;
Canada {\textendash} Natural Sciences and Engineering Research Council of Canada,
Calcul Qu{\'e}bec, Compute Ontario, Canada Foundation for Innovation, WestGrid, and Compute Canada;
Denmark {\textendash} Villum Fonden and Carlsberg Foundation;
New Zealand {\textendash} Marsden Fund;
Japan {\textendash} Japan Society for Promotion of Science (JSPS)
and Institute for Global Prominent Research (IGPR) of Chiba University;
Korea {\textendash} National Research Foundation of Korea (NRF);
Switzerland {\textendash} Swiss National Science Foundation (SNSF);
United Kingdom {\textendash} Department of Physics, University of Oxford.

\end{document}